**Approaches to studying virus pangenome variation graphs**

**Author information**
Tim Downing[1,2] [1] Dublin City University, Dublin, Ireland; [2] Pirbright Institute, Surrey, UK.
Email: Tim.Downing@pirbright.ac.uk Orcid: 0000-0002-8385-6730.

**Abstract**

Pangenome variation graphs (PVGs) allow for the representation of genetic diversity in a more nuanced way than traditional reference-based approaches. Here we focus on how PVGs are a powerful tool for studying genetic variation in viruses, offering insights into the complexities of viral quasispecies, mutation rates, and population dynamics. PVGs originated in human genomics and hold great promise for viral genomics. Previous work has been constrained by small sample sizes and gene-centric methods, PVGs enable a more comprehensive approach to studying viral diversity. Large viral genome collections should be used to make PVGs, which offer significant advantages: we outline accessible tools to achieve this. This spans PVG construction, PVG file formats, PVG manipulation and analysis, PVG visualisation, measuring PVG openness, and mapping reads to PVGs. Additionally, the development of PVG-specific formats for mutation representation and personalised PVGs that reflect specific research questions will further enhance PVG applications. Challenges remain, particularly in managing nested variants, optimising error detection, optimising k-mer/minimizer-based approaches for AT-rich genomes, incorporating long read sequencing data, and scalable visualisation approaches. Nevertheless, PVGs offer a new opportunities for viral population genomics, and a testing ground for tool development prior to application to larger eukaryotic genomes. These advances will enable more accurate and comprehensive detection of viral mutations, contributing to a deeper understanding of viral evolution and genotype-phenotype associations.

**Keywords:** genomics, pangenomics, virus, genome evolution, pangenome graph. [5 keywords]

## 1 What are pangenome variation graphs?

Pangenomes facilitate the investigation of broader patterns across a sample collection or lineage. Traditionally, a pangenome typically refers to the set of all genes in a sample collection (first coined by Sigaux in 2000 [1]). This could also be considered as the number of genetic elements [2]. Recently, gene-based pangenomes have been supplanted by more highly resolved nucleotide-level pangenomes, where each base can act as a genetic element of interest. These are called pangenome variation graph (PVG) or sometimes pangenome graphs: the former is used here.

A PVG is a graphical representation of the set of all mutations in a sample collection. PVGs can be compressed to allow faster analyses. Graphs can be used to represent genomic data in de Bruijn graph (DBG), string-based graph and other formats. Such graphs consist of nodes (or vertices), which represent sequences of varying lengths, and edges then link these nodes if they come from the same underlying sequence. We can then traverse through the PVG along the “path” associated with a single sample (or haplotype). Paths in the graph reflect the nodes observed in one or more samples in a specific order, and a traversal like this can be referred to as a “walk” (Table 1). For example, walking along a DBG starts with a single path that may split and merge at mutant sites that create bifurcating paths. These splits are termed bubbles that indicate alternate genetic sequences that split before merging later [3]. Superbubbles are an extension of this where multiple paths split, meander, and connect later (all at the same node). The diversity of these bubbles and superbubbles is a function of the input data [3].

## 2 Why should we use pangenome variation graphs?

Traditional methods to characterise samples map read libraries to existing linear reference genomes. A reference genome is a haploid sequence that does not contain any diversity, and thus is prone to the reference allele bias when reads are mapped to it. This is because a read will not map to the correct reference genome region if the number of nucleotide differences is high, resulting in missed mutations. Neither using a consensus reference sequence nor using a multi-sample linear reference addresses these limitations in viruses [4-5]. Mutation detection power is a function of genetic distance of the sample to the reference, resulting in impaired mutation detection for samples distantly related to the reference. As a result, accuracy decreases for divergent samples that align poorly to a single reference genome [2]. This is particularly limiting when examining reassortment or novel specimens. Such reference bias is a known limitation and means that variation at hypervariable loci was more difficult to resolve [6]. As a result, novel mutations, structural variants (SVs) and copy number variants (CNVs) may be missed [7]. Novel pathogen isolates genetically distinct from known references will be characterised less accurately [8-9]. Such missed or misinterpreted mutations result in imprecise inference of phylogenetic relationships and the molecular bases of phenotype changes [10-12].

**Table 1. Common terms in pangenome variation graph analysis.**

| Term | Definition |
|---|---|
| **Pangenome Variation Graph (PVG)** | A graphical representation of all mutations across a sample collection. |
| **De Bruijn Graph (DBG)** | A type of graph representing sequences by overlapping k-mers, where splits represent mutations. |
| **Bubble** | A PVG structure where a path splits and rejoins, indicating alternate genetic sequences. |
| **Superbubble** | A bubble structure where multiple paths diverge and reconverge, indicating complex mutations. |
| **Walk** | A path through a graph corresponding to the sequence of a single sample or haplotype. |
| **Sequence graph** | A single genome extracted from a graph as a connected set of nodes (a walk), often bidirected. |
| **Genome graph** | A path through a graph representing a full genome, factoring in gene orientation and annotation. |
| **Graph Burrows-Wheeler Transform (GBWT)** | A compressed index structure representing many paths through a graph for efficient mapping. |
| **Graphical Fragment / Assembly (GFA)** | A generic interpretable format for representing DBGs. |
| **Pairwise Alignment Format (PAF)** | A file format for recording pairwise alignments. |
| **Graph Alignment Format (GAF)** | A generalization of PAF format for use with PVGs in GFA. |
| **.vg format PVG** | A bidirectional graph for representing PVGs containing HashGraph or PackedGraph data. |
| **.pg format PVG** | A memory-efficient, static version of a PVG with PackedGraph data. |
| **.xg format PVG** | A static index of a PVG's paths, nodes and edges for queries. |
| **Generalized Compressed Suffix Array (GCSA)** | An index of suffixes in a graph used for efficient sequence location queries. |
| **Snarl** | A subgraph representing allelic or structural differences (like a bubble) between paths. |
| **Minimizer** | A subsequence of a k-mer representing for a sequence region; used for fast mapping. |
| **Graph Alignment/Map (GAM)** | A graph-based representation of read alignments on a PVG, equivalent to BAM/SAM files. |
| **PACK file** | The read support for candidate mutations in an augmented PVG. |
| **Surjection** | The conversion of a GAM file into a BAM file. |

Mapping reads to multiple linear reference genomes is one possible solution; however, this approach underperforms compared to using PVGs, which enable higher read mapping efficiency [13]. Another potential solution is *de novo* genome assembly, which assists in profiling novel specimens in principle. In practice, this approach is inadequate: errors in reference genomes, AT/GC content biases, host and vector contamination, short contigs, and repetitiveness can result in incomplete and biased genomes [14-15]. Moreover, contigs often need to be aligned to existing databases either as sequences or k-mers [16-17], and many dsDNA viruses require target enrichment or long-range or tiling PCR amplification, which require known reference genome(s). Although long read technologies are helping to address this genome assembly issue thanks to new approaches yielding lower error rates, the cost remains high. In addition, there are substantial numbers of high-quality genome assemblies and short read libraries for many important viruses, which can be examined effectively using PVG-based approaches. In this way, PVGs represent a cheaper approach to profile large number of published genomes and short read libraries.

A further consideration is linear reference genome quality. Some virus genomes lack high resolution of complex repeat structures, terminal inverted repeats (TIRs), regions with secondary structures, and polynucleotide tracts. This affects mutation detection at such regions, resulting in false positive and missed changes. PVGs reduce this bias by combining multiple genomes into one graph, thus allowing reads to map to paths that more closely reflect their true structures.

PVGs integrate genetic variation from an entire sample collection, capturing diversity that better reflects wider populations [6]. By representing alleles as distinct paths within the PVG, PVGs enable improved read mapping compared to traditional methods (Fig. 1). These paths encode various genomic features, such as alleles, haplotypes, alignments, and alternate sequence paths that correspond to genetic variation [3].

**Fig 1**. Pangenome variation graph construction and interpretation.

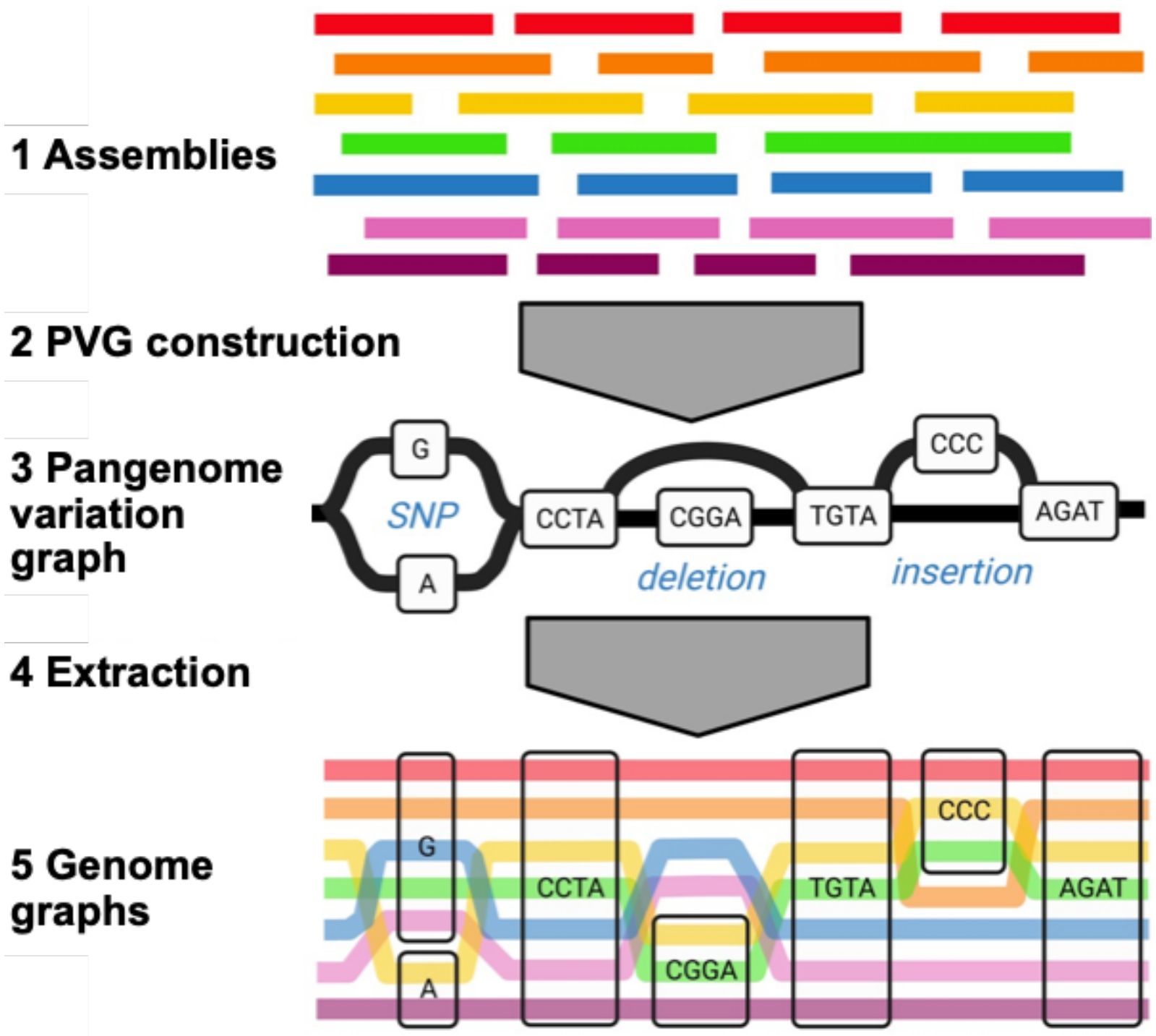

A set of assemblies (top) for seven different samples (represented by the colours) can be used to construct a variation graph (middle) to include all variations seen in this collection, such as SNPs or deletions or insertions. Not all samples possess each of these mutations, so the assembly genome graphs per sample (bottom) depict differing paths.

PVGs can be considered as DBGs where shared sequences are collapsed into single conserved nodes [6]. Rare nodes, which may represent sequencing errors, are typically removed during graph construction to ensure accuracy and allow easier manipulation of complex graph. For instance, the median error rate at bases generated by older Illumina MiSeq sequencing is ~0.5% [18]. A sample's genome extracted from a PVG is called a sequence graph: a single path indicated by a set of connected nodes that can be read in either forwards or reverse directions [19] (i.e, it is bidirected). PVGs are more easily considered as directed (i.e., each path has one interpretable directed), based on their functional annotation and CDS orientation. A genome graph is a sequence extracted from a PVG that is a set nodes in the PVG representing a whole genome, which has a directional component.

The primary advantage of using PVGs is a higher resolution of indels, structural variation (SV), including copy number variants (CNVs), small indels, and intrasample diversity. SVs include translocations, inversions, larger mutations and typically have lengths >50 bp, and small indels are <50 bp. This benefit from graphical formats is especially important when assessing paired-end short reads, where the total fragment size may limit power to detect large SVs, particularly compound SVs [7]. For instance, reads whose maximum fragment length is 900 bp cannot properly resolve SVs with lengths >900 bp because there would be no reads that fully span the SV length, rendering the exact length of the SV unclear. Although long reads possess better SV resolving power, this inferential capacity is still a function of read length that can be improved by PVGs. A secondary advantage of graphical formats is computationally scalable methods [20], allowing more thorough analyses [2]. As a consequence, PVGs have better sensitivity and accuracy [4,21-22].

Examples of this stem from work on humans. For example, SVs uniquely detectable using PVG-based methods has been found in rare human disease [23], GWAS in humans [24], in livestock [25], and in plants [26]. Over two-thirds of SVs were missed due to the usage of short reads mapped to a single reference genome in humans [27]. Moreover, PVGs found an average 64,000 additional mutations per human sample because 10% more reads were perfectly mapped to the PVG [28]: these hard-to-resolve regions/mutations are often strongly associated with phenotypic changes [29].

Consequently, PVGs are effective approaches for understanding genetic variation at the subspecies level, such as populations or clades, or any genetically related grouping lacking substantial structural genomic changes. Highly divergent samples or collections will prevent a PVG being formed properly: an example of this would be examining all African swine fever virus (ASFV) genomes, whose plastic genomes form tandem gene arrays with high CNV levels. Incomplete PVG possess low linearity and high levels of structural changes, such as bubbles and nested paths. An unanswered challenge in this area is to quantify how diverse informative PVGs can be.

**3 Pangenome variation graph construction**

A large number of PVG construction tools have been developed, applying the optimal approach of graph construction from genome assemblies (or sometimes reads) rather than from compiled mutation files, such as VCFs [7] (Table 2). These mutation files are insufficient because they were originally created based on comparisons of samples to a linear reference genome. As a result, they encode information specific to that comparison alone, omitting population-wide data associated with complex mutations.

Chief among the PVG construction tools is PGGB [30], which creates PVGs using three main steps: genome-wide pairwise alignment, PVG induction, and PVG refinement. The alignment stage, implemented by wfmash [31], examines 256 bp segments within regions of 1–25 Kb to cluster sequences sharing at least 90% similarity. This alignment approach is optimised so that pairwise alignment speed is efficient as the number of comparisons increases to a high level. The induction phase uses Seqwish, which constructs a PVG from the aligned sequences through an intermediate tree-based process [9]. Refinement is handled by Smoothxg [32], which splits the graph into blocks on which local realignment is completed to make partial order alignment (POA) graphs. Smoothxg is dependent on the input parameters, such as k-mer length (e.g 19 bp) and window size (e.g 27 bp). PGGB performs well across diverse datasets, including eukaryotic, prokaryotic, and viral genomes, and is available as part of the Nextflow nf-core [33]. It also generates extensive data exploration and visualisation options. However, its use of an all-to-all alignment approach means that scaling to large sample numbers can be challenging.

Another widely used tool is Minigraph, which can both construct PVGs and map reads to them [34]. Minigraph requires an initial PVG or reference genome and iteratively maps each sequence to the PVG, gradually reshaping it to incorporate observed diversity. It achieves this by identifying co-linear minimizer chains using Minimap2 [35], generating linear chains that are iteratively connected and incorporated into the PVG based on mapping quality and length. The resulting graph structure depends on the input order of the genome assemblies analysed. Minigraph was originally designed for large vertebrate genomes. Related to this is the Minigraph-Cactus pipeline [36], which combines Progressive Cactus alignment approaches [37] with Minigraph [34]. This pipeline was developed for larger genomes and is part of the Cactus software package.

A related tool, Scaffold Graph Toolkit [38], enables PVG construction in the form of scaffold graphs in formats such as FASTQ or GFA. It supports associated analyses and interactive visualisation through Cytoscape. Input data can include FASTQ, FASTA, SAM, BAM, or contig file formats.

An important consideration for PVG construction is scalability, which has been enhanced by indexing a PVG as a graph Burrows-Wheeler Transform (GBWT) by ODGI [41] and Giraffe [20] without the need search across the graph itself. Additionally, indexes like the r-index scale better with large datasets [42]. Nonetheless, scalability is still a function of virus genomic diversity, length and repetitiveness. Existing approaches alignment sequence fragments to one another to iteratively build a PVG that can be pruned [30,36,38]. However, viruses with high diversity will create a complex PVG with numerous paths, nested variants, bubbles and superbubbles. Thus, considerable thought must be given to the taxonomic unit of interest (population, subclade, clade, serotype or species) such that the PVG can be interpreted effectively.

Pangenome graph file formats are also an essential consideration, with Graphical Fragment Assembly (GFA) being the primary format (https://github.com/GFA-spec/GFA-spec). However, GFA currently lacks coordinate support. Reference GFA (rGFA) [34] includes reference paths in the form of walks (i.e sets of connected nodes) but may not fully accommodate the representation of complex mutations. There are two commonly used GFA versions: GFA1.0, used in PGGB-related methods, and GFA1.1, associated with Minigraph workflows. The third, GFA2 is less widely used at present (https://gfa-spec.github.io/GFA-spec/). Methods for each GFA type are typically specific to its specification, with limited interoperability between GFA1.0 and GFA1.1. Tools like PGGB generate pairwise alignment files in the PAF format [30], which has been extended to the Graph Alignment Format (GAF) for PVGs in GFA format. GFA files can be converted to GAF or PAF with minichain [43], where GAF generalises the text-based PAF format. Lastly, a PVG Format (PGVF), derived from GFA1.0, for graph-to-graph comparisons has been proposed but remains unsupported this far (https://github.com/pangenome/pgvf-spec).

## 4 Pangenome variation graph analysis, manipulation and visualisation

PVG analysis and visualisation tools are essential for deciphering both simple and complex structural variations (SVs). These tools extend and integrate with PVG construction tools but often have significant back-end complexity or lack command-line interfaces. Among the most prominent is the optimized dynamic genome/graph implementation (ODGI), which enables computationally efficient PVG examination and is an indispensable tool for PVG analysis with (currently) 44 functions [41]. ODGI provides extensive functionality, including PVG manipulation, node trimming, sorting, exporting, and various refinements, along with visualisation capabilities. By default, ODGI works with GFA1.0 format but can be toggled via Minigraph-Cactus to support GFA1.1 using the Variation Graph (VG) toolkit's convert function. ODGI works with PVGs in .og format and has a partial Python equivalent called mygfa (https://github.com/cucapra/pollen/tree/main/mygfa).

Numerous other PVG analysis tools complement ODGI. Gretl (https://github.com/MoinSebi/gretl) provides rapid summaries of GFA statistics, core metrics, and path similarity metrics, including rates per node and sequence amounts. Gfatools [44] enables subgraph extraction from a GFA and conversion of GFA files to FASTA or BED format. Gaftools (https://github.com/marschall-lab/gaftools) offers a wide range of GAF-related functions, including GFA indexing to GAF, GAF viewing, subsetting, alignment sorting, realignment, subsequence extraction from node paths, PVG ordering, phasing using TSV data, and numerical summarisation of GAF files. Gfakluge can sort PVGs in GFA format, extract FASTA sequences from GFA, convert between GFA1.0 and GFA1.1, and merge GFA pairs. Similarly, Gfapy can perform conversions between GFA1.0 and GFA1.1 [45]. Impg allows the extraction of a PVG region based on a region of interest in a particular sample/path (https://github.com/pangenome/impg). Another tool, Gafpack, computes coverage metrics for GAF files based on GFAs (https://github.com/pangenome/gafpack). Chrom_mini_graph create a coloured minimizer PVG based on FASTQ queries (https://github.com/gaojunxuan/chrom_mini_graph). Gfalace usefully allows the merging of different PVGs (https://github.com/pangenome/gfalace).

In terms of PVG quality, there are four main options. The first is the pangenome graph evaluator (PGGE, https://github.com/pangenome/pgge) assesses PVG accuracy and optimises PVG construction across various approaches. A second is the pangenome single copy and universal k-mer counter (PG-SCUnK), which assesses PVG quality using single-copy and universal k-mers based on the principle that single-copy k-mers are orthologous across samples (https://github.com/cumtr/PG-SCUnK/, [46]). A third is ODGI validate, which compares the PVG topology to its constituent paths [41]. And fourthly, VG validate can examine PVG nodes, node structures, path connectedness across edges, and also the compatibility of a GAM file with a PVG [22].

Despite the availability of analysis tools, effective PVG visualisation tools for non-human organisms remain limited. One of the best options is SequenceTubeMap [47], which visualises PVGs in .pg or .vg formats via a web interface or Docker installation. It supports graph alignment/map (GAM) file and/or haplotype data, as well as BED file annotations. However, SequenceTubeMap imposes a 5 MB upload limit, which may require down-sampling certain GAM files. Other visualisation tools include ODGI's view function, gfaestus for 2D visualisation (https://github.com/chfi/gfaestus), the Variation Graph (VG) toolkit's view function [22], pvg (https://github.com/w-gao/pgv), the Assembly Graph Browser (AGB) [48], Waragraph (https://github.com/chfi/waragraph) for generating 1D and 2D plots, Bandage which has been effective for graphical visualisation for some time [49], and gfaviz [50], which creates 2D plots, scaffold graphs, alignment graphs, and PVG images. A newer option is wally, which visualises the mutations in mapped short reads, long reads, contigs and PVGs using BAM (or CRAM) input (https://github.com/tobiasrausch/wally) [51].

## 5 Pangenome variation graph read mapping & mutation detection

A wide variety of tools exist for constructing and analysing PVGs based on multiple sequence alignment (MSA) and read mapping (Table 2). Notably, read mapping to a PVG demonstrates computational efficiency comparable to mapping against linear references [2]. Earlier methods focused on mapping short reads to de Bruijn graphs (DBGs), but modern tools enhance efficiency using k-mer-based strategies, suffix trees, and BWTs. VG's map function (VG-MAP) mapped reads to a PVG in the same manner as mapping to a linear reference where a read could map to the range of haplotypes in the PVG [22]. However, this has been superseded by VG's giraffe function (Giraffe). To map with Giraffe, VG uses a BWT for efficient PVG indexing and read mapping using a POA to create GBWT and GBZ indices [7]. Giraffe maps reads to the combination of alleles across haplotypes, using minimizer and genetic distance data as additional input. Consequently, Giraffe maps reads to synthetic haplotypes absent in the original PVG. For this reason, Giraffe is much more effective that VG-MAP and VG-MPMAP - in the latter, reads are mapped to multiple paths [20]. Effective SV detection tools using graph include Paragraph [52], VG [22] and Graphtyper2 [53]. Paragraph and GraphTyper2 use different approaches to re-map reads at known mutations in the reference graph to improve the original detection based on a linear reference [52-53]. Lastly, PanGenie uses a k-mer-based approach to support genotyping, and high good specificity in detecting mutations and sub-consensus changes in closely related specimens [54].

**Table 2. Pangenome variation graph mapping tools suitable for virus PVG analyses.**

| Name | Description | Ref. |
|---|---|---|
| GraphTyper2 | Uses short reads to genotyping mutations with a full PVG | [53] |
| Paragraph | Uses short reads to genotyping structural variants with a local PVG | [52] |
| PanGenie | Uses short reads' k-mers to genotype with haplotype-resolved PVG | [54] |
| Giraffe | Map short reads to allelic variation | [20] |
| VG-MAP | Map short reads to known haplotypes | [7] |
| VG-MPMAP | Map short reads to multiple paths | [7] |
| V-MAP (Variant Map) | Map short or long reads to a PVG | [55] |
| Minigraph | Map short or long reads to a PVG & can construct PVGs too | [34] |
| GraphAligner | Map long reads to a PVG | [56] |
| PanAligner | Map long reads to a PVG | [57] |

In addition, some tools can now map long reads to reference graphs, such as Giraffe [20], GraphAligner [56], Minigraph [34], V-Map (Variant Map) [55] and PanAligner [57]. GraphAligner maps long reads and may be better thanks to banded sequence alignment, whereas V-Map tolerates both short and long reads [55-56]. Other options do exist, but these are either old or are designed with human data in mind.

Like ODGI above, PVG analysis and mapping needs to include the VG toolkit [7]. VG can integrate information from VCFs, assemblies and create PVGs representing the full spectrum of known diversity in a sample collection. Consequently, VG is an analysis tool as well as one for read mapping. VG has a range of file conversion capabilities, as well as ones to view PVG components. Although VG's construct function can be used to make a PVG based on VCF data, this is not advised if the mutation data comes from linear reference mapping. Gfautil (https://crates.io/crates/gfautil) and VG's deconstruct function allow conversion in the other direction (from GFA to VCF).

For PVG read mapping, there are more complex file format options that are akin to SAMtools (REF) faidx indexing. VG's convert function can convert PVGs into the bidirectional .vg (HashGraph or PackedGraph) format, allowing the PVG to be modified by VG (Fig. 2). Optionally, long nodes can be reduced in size using VG's mod function for large PVGs. To create a compressed haplotype index, VG's gbwt function can convert GFA files into GBZ or GBWT format. Another common PVG file format is .xg (XG), which is a static sequence-free PVG index for querying. The PVG .xg format can be converted into .pg (Packed Graph) format with VG's convert function. PVGs can also be indexed in generalised compressed suffix array index (GCSA) format, which contains a suffix array of sequence locations on the PVG [58].

**Fig 2. Pangenome variation graph file format and tool functions.**

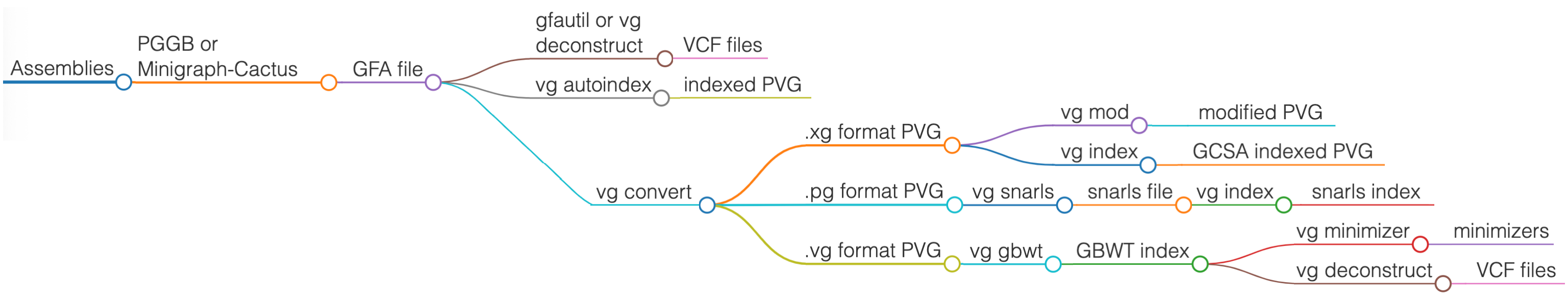

Fig. 2 legend: A set of assemblies (top) can be converted into GFA format, and from then into VCF files by VG deconstruct or Gfautil, and index with VG autoindex. The GFA file can be converted by VG's convert function into .vg or .pg or .xg formats. The PVG in .xg format can be modified by VG mod, and VG's index function can create a generalised compressed suffix array index (GCSA) from it. The .pg format PVG can be used to make a snarls file using VG snarls, then a snarls index from that using VG index. The .xg format PVG can be used to make a GBWT index using VG's gbwt function. This index can be used to make VCF files using VG deconstruct and also minimizers using VG minimizer (if given a set of genetic distances as additional input).

The next step is PVG pruning: the removal of rare or discontinuous nodes. For small genome sizes such as those of viruses, this may not be needed, but for large or complex datasets this is crucial. Following this, VG's snarls function can be used to create a snarl index from the .pg format PVG [3] - a snarl is a subgraph representing allelic or structural difference between paths, sometimes also called a bubble.

VG's deconstruct function can take the PVG in .xg format and the GBWT index to create a VCF of the PVG's diversity. VG's autoindex function is useful because it automatically detects the appropriate indexing approach based on the input data. Lastly, VG's minimizer function can use the GBWT index (and a set genetic distances) to create a set of minimizers to support other analyses.

In relation to read mapping file formats, GAM is analogous to sequence alignment/map (SAM) and binary alignment/map (BAM) [22]: they encode the coordinates of the reads' matches on the PVG. GAM files are also equivalent to GAF files. VG's stat function can generate mapping metrics on a GAM file. VG augment reads in a GAM file and the PVG in .pg format to create an augmented PVG: the latter integrates the mutations in the GAM file with that PVG. VG pack uses this augmented PVG to make a PACK file, generating read support for each candidate mutation. VG call can make a VCF file for that sample's reads using this PACK file, the augmented PVG, and a snarls file. In parallel, VG depth takes in the .vg format PVG and the PACK file to create a file containing the read depth across the PVG. VG can surject GAM files into BAM format for processing with widely used tools such as Freebayes [59], DeepVariant [60], BCFtools [61] and GATK [62].

## 6 Pangenome variation graph openness

The number of samples required to create a representative PVG is a specific to the research question and sample collection involved. This can be assisted using Heap's Law to model the rates of new mutations as the number of samples included increases [63]. If we continue to find new mutations (on average) as more samples are added to the PVG, we can say the PVG is open. If we no longer find new mutations, then it is considered a closed PVG. The number of new mutations in *N* genomes sharing *n* mutations can be estimated from the power-law regression $n=\kappa N^{gamma}$ where κ is a scalar [80] (Fig. 3A). The rate of new mutation (Δn) can be calculated from $\Delta n=kN^{-alpha}$ where k is another scalar, and an open PVG has alpha <1 and a closed one alpha >1 [64] (Fig. 3B).

**Fig. 3**. Modelling PVG openness.

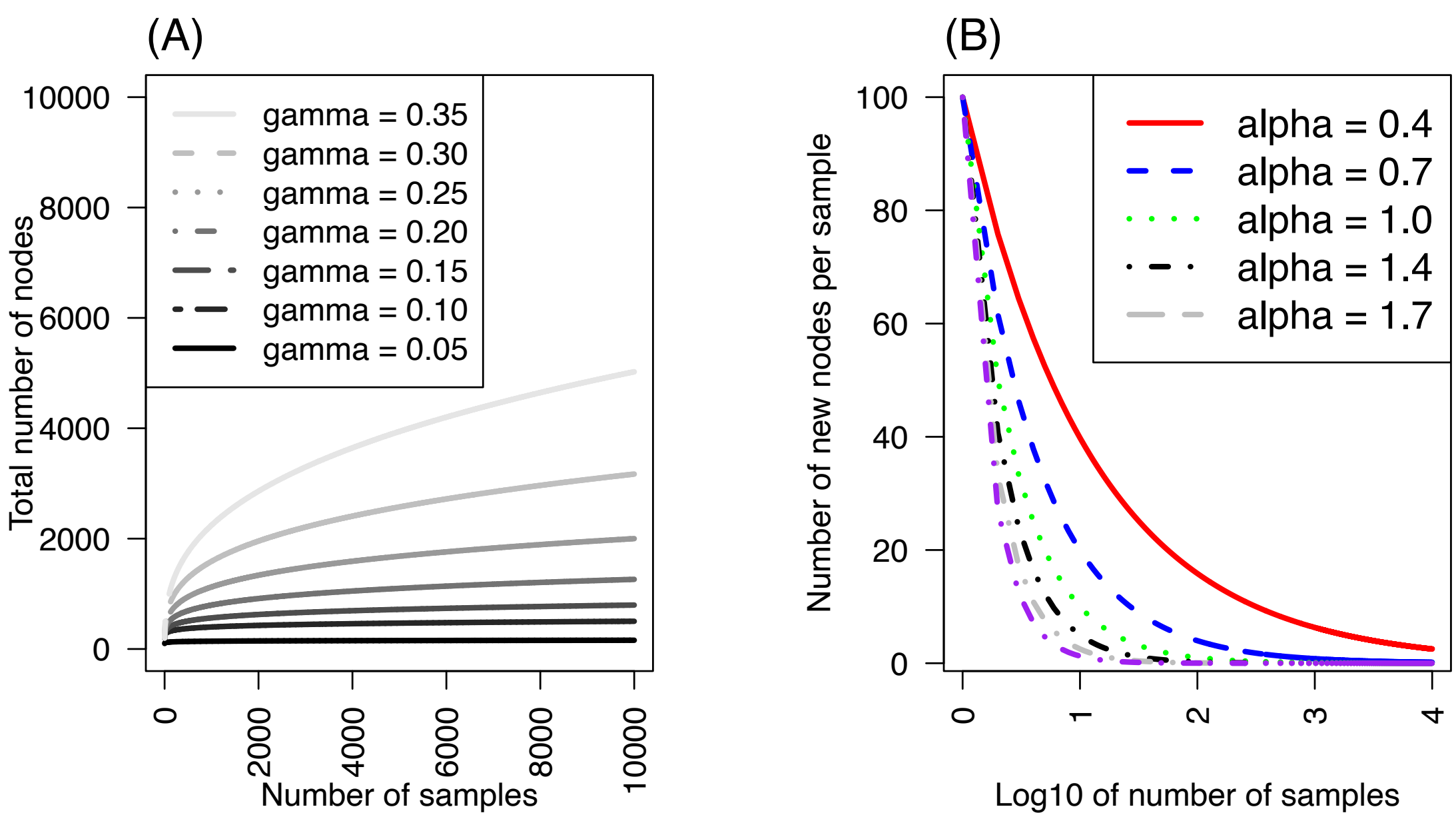

Fig. 3 legend: The (A) total number of nodes, and (B) number of new nodes per sample as the number of samples (x-axes) increases. (A) The grey lines show differing values of gamma such that the total number of nodes in N samples is $\kappa N^{gamma}$ where κ=100. (B) The coloured lines show varying levels of alpha such that the number of new nodes is $kN^{-alpha}$ where k=100.

There are three main approaches for estimating PVG openness. This first is Pangrowth, which estimates this based on the rate of increase in number of unique k-mers as more genomes are added to the PVG [65]. The second is Panacus, which uses rate of growth in nodes in the PVG as the genome number increases [66]. These are important because they allow computation of the sample size needed to reflect the population's broader diversity based on the sample taken (Fig. 3A). As the relative rate of new k-mers or nodes becomes asymptotically close to zero, additional samples may not novel variation that informs on genomic or evolutionary questions in the dataset. Such a PVG would be closed with alpha >1 (Fig. 3B).

In a gene-based pangenome, the core genome is composed of genes shared by (usually) >99% samples, and the remaining genes at lower frequencies (typically <95%) represent the accessory genome [63]. The precise thresholds used here (>99%, >95%, <95%) for core, accessory and other strata depend on the genomes being examined. OGDI's heaps function computes the number of bases in a PVG as the number of genome in the PVG increases and tells us more about the PVG-based terms for core and accessory pangenomes [41]. Firstly, the shared PVG from ODGI heaps reflects the sum of bases at nodes found in all genomes, analogous to the core genome. Secondly, the sum of bases at any node is the complete PVG, i.e. every single base in the PVG. Thirdly, the sum of nodes in 50% of samples indicates the median PVG size, which may estimate the median genome size in the sample set.

## 7 Viral haplotype reconstruction from pangenome variation graphs

Viral haplotype reconstruction is closely linked to PVGs: both processes aim to represent the genetic variation within a population of viral genomes. The frequencies of major and minor haplotypes can vary across different samples and genomic regions. Viral haplotype reconstruction is a form of strain-aware metagenome assembly, and it is particularly complex due to the varying viral effective population sizes and mutation rates, which lead to significant differences in within-host sequence diversity [67]. For viruses with segmented genomes, an additional layer of complexity arises from the difficulty in resolving the linkage between segments. These linkages are often challenging and may not be fully resolvable in some cases. Current haplotyping methods, which typically rely on de novo assembly or read mapping to a reference genome, show varied outcomes and can struggle when applied to highly variable genomic data [67]. Furthermore, the depth of sequencing reads is an important factor in detecting rare minor alleles. Most read mapping tools have maximum mismatch rates, which are set based on the error rates of the sequencing technology. These constraints can limit the detection of rare variants, which may be crucial for understanding viral diversity within a population.

Viral haplotypes can be reconstructed from the paths in the PVG, where the inter-node coverage correlation plays a significant role. In most cases, these resolved viral haplotypes correspond to paths in the PVG, and since they are largely composed of long unitigs, most of the nodes are relatively stable with few mutations. However, regions of the genome with low complexity can complicate this process, leading to cycles and bubbles in the potential paths. Typically, directed acyclic graphs (DAGs) are sufficient when representing homogeneous viral genomes because they assume a start and an end. But complex ones might need bidirected graphs, which model contig orientation and SVs more accurately. One approach to resolving these complexities is to first build a contig PVG using initial *de novo* assembly contigs and sequencing read libraries. From there, the relative change in allele frequencies across alternate contig paths is minimized [68]. This method uses the coverage of mapped reads to help determine the relative frequencies of each haplotype across the contigs, which is essential for resolving the viral haplotypes. Once the haplotypes are resolved, they constitute individual genome graphs that, when aggregated, form the overall PVG.

Other graphical representations of viral quasispecies have been developed to handle mixed sample metagenomes. Tools like SAVAGE [12], Virus-VG [39], and VG-Flow [68], which use flow variation graphs, offer solutions for representing haplotype abundance and further improving the accuracy of viral haplotype reconstruction in complex viral populations. For all approaches, read error correction before assembly can be beneficial to improve minor allele detection. Tools that perform this correction assume that sequencing errors are rare k-mers in a graph and not rare haplotypes [19]. This approach is particularly useful because PVGs, due to their structure, may highlight rare haplotypes that might otherwise be missed.

## 8 Potential for pangenome variation graphs applied to viral genomes

PVGs originated in human genomics and there are significant opportunities for their use to illuminate viral genetic variation. No PVG analyses on large viral sample collections have been conducted yet. Pangenomic approaches has been applied to viruses, including 59 hepatitis B read libraries [13], eight *Pithoviridae* genomes [69], 160 Ebolavirus genomes [70], 42 ASFV genomes [71], 46 ASFV genomes [72], and 36 chlorovirus genomes [73]. Excluding hepatitis B and Ebolavirus, these studies have been constrained by gene-centric methods and small sample sizes (n<50).

Animal and plant pathogenic viruses typically have genome lengths <30 Kb [74], potentially rendering PVG-based approaches less informative. The genome sizes of dsDNA viruses are usually larger (4.7 Kb to 2.47 Mb), whereas ssDNA viruses are usually smaller (range: 0.86-10.9 Kb) [74]. PVG-style approaches could be relatively accessible for the >715 DNA and RNA viral genera whose genome is a single molecule, which include all dsDNA and most ssDNA viruses. 46 additional genera have a segmented genome contained within a single particle, and a further 45 have multipartite genomes where each segment is packaged in a separate particle. For the latter 91 genera, PVG construction is more complex, but can avail of approaches from vertebrates.

A paradigm in this area is PVG investigation of the 3.2 Kb hepatitis B PVG [13]. This genome is unusual: it is a small, partially dsDNA and circular, and yet can replicate independently in hosts in part thanks to four partially overlapping ORFs. Existing linear hepatitis B reference genomes showed wide heterogeneity of mapping outcomes for real and simulated read libraries, and that no such reference performed as well as a representative PVG [13]. Additionally, this accuracy was a function of genetic divergence such that PVGs contributed more information for samples with higher diversity [13]. This was particularly acute at more variable regions, which may be associated with the >8% of reads that did not map to linear references but did to a PVG [13]. Although there were instances where the PVG vs linear mapping rate difference was small, predicting this in advance of any analysis of an uncharacterised sample would be challenging.

As an example of where PVGs could be useful, African swine fever virus (ASFV) is paramount due to its dynamic genome. ASFV genomes are dsDNA molecules of 170.1 to 193.9 Kb lengths with 150-167 ORFs on both strands [71-72,75]. ASFV is in the nucleocytoplasmic large dsDNA virus (NCLDV) superfamily and the Asfarvirus family. Notably, the ASFV genome contains nested genes within larger ORFs that were embedded in polycistronic arrays, rather than a monocistronic one [76]. Across all genotypes but with high sampling of genotypes I and II, ASFV has an open pangenome suggesting that broader population ASFV sampling will identify additional novel genes. This is based one on study of 42 genomes that had an open pangenome (alpha=0.62) with 102 core and 168 accessory genes [71], which was extended by a study of 46 genomes that had 86 core and 324 accessory gene clusters [72]. The variation in these estimates is understandable given ASFV's high diversity and the comparatively small sample sizes in each investigation.

Based on the information above and with a view to identifying the most straightforward path for viral PVG studies, suggested key tools are listed to enable researchers new to PVGs to navigate this complex area (Table 3). The first four stages (PVG creation, analysis, visualisation and openness) refer to PVGs themselves, whereas the final stage of read mapping is an application of PVGs to mutation detection. New PVG tools are emerging rapidly [77], and the following website maintains an updated list: https://github.com/colindaven/awesome-pangenomes.

## 9 Future prospects

PVGs should serve as the cornerstone for many population genomics studies, including multi-omics ones [78]. While PVG analysis and visualisation is an emerging field, significant software development and formalisation are needed to reduce reliance on bespoke scripts. One limitation of current PVG approaches is the continued use of linear VCF formats to represent mutations. This practice limits the output's utility, suggesting the need for the development of a dedicated PVG-based format for VCF files [36]. Moreover, there is a growing need for enhanced capacity to handle nested variants and optimise error detection across large sample collections and diverse sequencing technologies. Linked to this is the opportunity to create a wider array of PVG visualisation options that are scalable, rapid, present layers of information effectively, and permit non-experts to view mutations of interest. The small size of viral genomes mean they offer an efficient sandbox for PVG visualisation development. These all would significantly improve the effectiveness of PVGs in advancing scientific investigations.

**Table 3. Suggested key tools for viral pangenome variation graph analyses.**

| Tool | Stage | Task | Ref. |
|---|---|---|---|
| PGGB | Creation | PVG construction | [9] |
| ODGI | Analysis | PVG analysis & manipulation | [41] |
| SequenceTubeMap | Visualisation | PVG visualisation | [47] |
| Panacus | PVG openness | PVG openness | [66] |
| Pangrowth | | PVG openness | [65] |
| ODGI heaps | | PVG size | [42] |
| VG | Read mapping | PVG indexing | [9] |
| Giraffe | | Short read mapping | [20] |
| GraphAligner | | Long read mapping | [70] |

One promising avenue to overcome some of these limitations is the use of k-mer-based methods. Such approaches are more efficient and scalable than alignment-based ones, and so they offer a powerful way to quantify diversity and rates of genetic change in large collections. At the same time, k-mer-based methods risk lower accuracy and sensitivity in detailed analyses such as detecting specific mutations due to shared k-mers and the value of k used [87]. For the latter, alignment-based approaches are more effective. Emerging k-mer tools, such as pankmer [80], enable reference-free k-mer-based analyses, while other methods index all k-mers in the input reads to construct the PVG [81]. These k-mer-based approaches can be computationally more efficient, making them more scalable [82]. An example is the PVG indexing method called Maximal Exact Match Ordered (MEMO) uses k-mers to allow rapid k-mer matching queries at a variety of scales [83]. As sequencing error rates continue to decline, these methods will also become more accurate [18].

A key challenge in PVG analysis is managing the sheer volume of mutations included in the graph, which can place significant strain on computational resources [84]. This has led to the development of reference PVGs [13,28,34]. However, the reference set of genomes used to create a PVG should be carefully selected to align with the specific scientific question being addressed. A promising solution is the use of "personalised" PVGs, where haplotypes are chosen based on patterns of k-mer count matching in read data. These personalised PVGs outperform those based on common variants or the entire sample set [85]. Ideally, these PVGs would be composed of hybrid assemblies generated from both short and long-read sequencing technologies. Fortunately, viruses present fewer computational and phasing challenges for this k-mer matching process, making them well-suited for these advanced approaches, including those that link existing methods together [86]. At the same time, many viral genomes are complex (e.g. circular) or may be AT-rich, thus requiring a different approach when considering "personalised" PVGs.

A final note is that PVGs may not be effective for genetically divergent novel samples without a closely related comparator genome. Fortunately, virus long Oxford Nanopore Technology (ONT) and Pacific Biosciences (PacBio) read sequencing error rates are continuing to decrease, suggesting that this may be a useful option. This can be integrated with PVG-based approaches: it is now possible to augment a PVG to incorporate a new sample's long read library (https://github.com/ldenti/palss). Moreover, genotyping may be enhanced by the incorporation of high-quality long read assemblies, for which PVGs are effective visualisation aids.

**Competing Interests Statement**

The author declares no competing interests.

**Acknowledgements**

The Pirbright Institute receives support through from UK Research and Innovation (UKRI) Biotechnology and Biological Sciences Research Council (BBSRC) grants BBS/E/PI/230002A, BBS/E/PI/230002B, BBS/E/PI/230002C and BBS/E/PI/23NB0003. I thank the anonymous reviewers for their constructive comments and feedback.

**References**

1. Sigaux F. Génome du cancer ou de la construction des cartes d'identité moléculaire des tumeurs. Bull Acad Natl Med. 2000;184(7):1441-7.
2. Eizenga JM, et al Pangenome Graphs. Annu Rev Genomics Hum Genet. 2020;21:139-162. doi: 10.1146/annurev-genom-120219-080406
3. Paten B, Eizenga JM, Rosen YM, Novak AM, Garrison E, Hickey G. Superbubbles, Ultrabubbles, and Cacti. J Comput Biol. 2018;25(7):649-663. doi: 10.1089/cmb.2017.0251.
4. Chen NC, Solomon B, Mun T, Iyer S, Langmead B. Reference flow: reducing reference bias using multiple population genomes. Genome Biol. 2021;22(1):8 doi: 10.1186/s13059-020-02229-3.
5. Chen NC, Paulin LF, Sedlazeck FJ, Koren S, Phillippy AM, Langmead B. Improved sequence mapping using a complete reference genome and lift-over. Nat Methods. 2024;21(1):41-49. doi: 10.1038/s41592-023-02069-6.
6. Rakocevic G, et al Fast and accurate genomic analyses using genome graphs. Nat. Genet. 2019;51:354–362.
7. Hickey G, Heller D, Monlong J, Sibbesen JA, Sirén J, Eizenga J, Dawson ET, Garrison E, Novak AM, Paten B. Genotyping structural variants in pangenome graphs using the vg toolkit. Genome Biol. 2020;21(1):35. doi: 10.1186/s13059-020-1941-7
8. Boehm E, Kronig I, Neher RA, Eckerle I, Vetter P, Kaiser L; Geneva Centre for Emerging Viral Diseases. Novel SARS-CoV-2 variants: the pandemics within the pandemic. Clin Microbiol Infect. 2021 27(8):1109-1117. doi: 10.1016/j.cmi.2021.05.022
9. Garrison E, Guarracino A. Unbiased pangenome graphs. Bioinformatics. 2023;39(1):btac743. doi: 10.1093/bioinformatics/btac743
10. Rick JA, Brock CD, Lewanski AL, Golcher-Benavides J, Wagner CE. Reference Genome Choice and Filtering Thresholds Jointly Influence Phylogenomic Analyses. Syst Biol. 2024;73(1):76-101. doi: 10.1093/sysbio/syad065
11. Valiente-Mullor C, et al. One is not enough: On the effects of reference genome for the mapping and subsequent analyses of short-reads. PLoS Comput Biol. 2021;17(1):e1008678. doi: 10.1371/journal.pcbi.1008678.
12. Baaijens JA, Aabidine AZE, Rivals E, Schönhuth A. De novo assembly of viral quasispecies using overlap graphs. Genome Res. 2017;27(5):835-848. doi: 10.1101/gr.215038.116
13. Duchen D, Clipman S, Vergara C, Thio CL, Thomas DL, Duggal P, Wojcik GL. A hepatitis B virus (HBV) sequence variation graph improves sequence alignment and sample-specific consensus sequence construction for genetic analysis of HBV. PLoS One. 2024;19(4):e0301069. doi: 10.1371/journal.pone.0301069

14. Alser M, et al. Technology dictates algorithms: recent developments in read alignment. Genome Biol. 2021;22(1):249. doi: 10.1186/s13059-021-02443-7
15. Lischer HEL, Shimizu KK. Reference-guided de novo assembly approach improves genome reconstruction for related species. BMC Bioinformatics. 2017;18(1):474. doi: 10.1186/s12859-017-1911-6
16. Bradley P, den Bakker HC, Rocha EPC, McVean G, Iqbal Z. Ultrafast search of all deposited bacterial and viral genomic data. Nat Biotechnol. 2019;37(2):152-159. doi: 10.1038/s41587-018-0010-1
17. Ondov BD, Treangen TJ, Melsted P, Mallonee AB, Bergman NH, Koren S, Phillippy AM. Mash: fast genome and metagenome distance estimation using MinHash. Genome Biol. 2016;17(1):132. doi: 10.1186/s13059-016-0997-x.
18. Stoler N, Nekrutenko A. Sequencing error profiles of Illumina sequencing instruments. NAR Genom Bioinform. 2021;3(1):lqab019. doi: 10.1093/nargab/lqab019
19. Baaijens JA, et al Computational graph pangenomics: a tutorial on data structures and their applications. Nat Comput 2022;21:81–108 doi: https://doi.org/10.1007/s11047-022-09882-6.
20. Siren, J. et al Genotyping common, large structural variations in 5,202 genomes using pangenomes, the Giraffe mapper, and the vg toolkit. bioRxiv 2020 https://doi.org/10.1101/2020.12.04.412486.
21. Garrison E, et al Variation graph toolkit improves read mapping by representing genetic variation in the reference. Nat Biotechnol. 2018;36(9):875-879. doi: 10.1038/nbt.4227
22. Novak AM, Garrison E, Paten B. A graph extension of the positional Burrows-Wheeler transform and its applications. Algorithms Mol Biol. 2017;12:18. doi: 10.1186/s13015-017-0109-9.
23. Groza C, et al Pangenome graphs improve the analysis of structural variants in rare genetic diseases. Nat. Commun. 2024;15:657.
24. Gupta PK. GWAS for genetics of complex quantitative traits: genome to pangenome and SNPs to SVs and k-mers. Bioessays 2021;43:e2100109.
25. Li R, et al A sheep pangenome reveals the spectrum of structural variations and their effects on tail phenotypes. Genome Res. 2023 33:463–477.
26. Gao L, et al The tomato pan-genome uncovers new genes and a rare allele regulating fruit flavor. Nat. Genet. 2019;51:1044–1051.
27. Zhao X, et al. Expectations and blind spots for structural variation detection from long-read assemblies and short-read genome sequencing technologies. Am J Hum Genet. 2021;108(5):919-928. doi: 10.1016/j.ajhg.2021.03.014. v
28. Liao WW, et al. A draft human pangenome reference. Nature. 2023;617(7960):312-324. doi: 10.1038/s41586-023-05896-x.
29. Haga IR, et al. Sequencing and Analysis of Lumpy Skin Disease Virus Whole Genomes Reveals a New Viral Subgroup in West and Central Africa. Viruses. 2024;16(4):557. doi: 10.3390/v16040557
30. Garrison E, et al. Building pangenome graphs. Nat Methods. 2024;21(11):2008-2012. doi: 10.1038/s41592-024-02430-3
31. Guarracino A, Mwaniki N, Marco-Sola S, Garrison E. wfmash: whole-chromosome pairwise alignment using the hierarchical wavefront algorithm. 2021 https://github.com/ekg/wfmash.
32. Garrison, E. et al. smoothxg: normalization of variation graphs with local partial order realignment. Zenodo 2022 https://doi.org/10.5281/zenodo.7239231
33. Heumos S, et al Cluster-efficient pangenome graph construction with nf-core/pangenome. Bioinformatics 2024;40(11):btae609. doi: 10.1093/bioinformatics/btae609.
34. Li H, Feng X, Chu C. The design and construction of reference pangenome graphs with minigraph. Genome Biol. 2020;21(1):265. doi: 10.1186/s13059-020-02168-z
35. Li H. Minimap2: pairwise alignment for nucleotide sequences. Bioinformatics. 2018 34(18):3094-3100. doi: 10.1093/bioinformatics/bty191.
36. Hickey G, et al Pangenome graph construction from genome alignments with Minigraph-Cactus. Nat Biotechnol. 2024;42(4):663-673. doi: 10.1038/s41587-023-01793-w.
37. Armstrong J, et al Progressive Cactus is a multiple-genome aligner for the thousand-genome era. Nature. 2020;587(7833):246-251. doi: 10.1038/s41586-020-2871-y.
38. Kunyavskaya O, Prjibelski AD. SGTK: a toolkit for visualization and assessment of scaffold graphs. Bioinformatics. 2019;35(13):2303-2305. doi: 10.1093/bioinformatics/bty956.

39. Baaijens JA, et al. Full-length de novo viral quasispecies assembly through variation graph construction. Bioinformatics 2019;35:5086–94.
40. Audano PA, et al Characterizing the major structural variant alleles of the human genome. Cell 2019;176:663–675.e19
41. Guarracino A, Heumos S, Nahnsen S, Prins P, Garrison E. ODGI: understanding pangenome graphs. Bioinformatics. 2022;38(13):3319–26. doi: 10.1093/bioinformatics/btac308.
42. Moeckel C, Mareboina M, Konnaris MA, Chan CSY, Mouratidis I, Montgomery A, Chantzi N, Pavlopoulos GA, Georgakopoulos-Soares I. A survey of k-mer methods and applications in bioinformatics. Comput Struct Biotechnol J. 2024;23:2289-2303. doi: 10.1016/j.csbj.2024.05.025
43. Chandra G, Gibney D, Jain C. Haplotype-aware sequence alignment to pangenome graphs. Genome Res. 2024;34(9):1265-1275. doi: 10.1101/gr.279143.124
44. Li, H. Gfatools: tools for manipulating sequence graphs in the GFA and rGFA formats. GitHub https://github.com/lh3/gfatools 2024.
45. Gonnella G, Kurtz S. GfaPy: a flexible and extensible software library for handling sequence graphs in Python. Bioinformatics 2017;btx398 https://doi.org/10.1093/bioinformatics/btx398
46. Tristan C, Sotiria M, Leonard AS, Hubert P. PG-SCUnK: measuring pangenome graph representativeness using single-copy and universal K-mers. BioRxiv 2025 doi: https://doi.org/10.1101/2025.04.03.646777
47. Beyer W, Novak AM, Hickey G, Chan J, Tan V, Paten B, Zerbino DR. Sequence tube maps: making graph genomes intuitive to commuters. Bioinformatics. 2019;35(24):5318-5320. doi: 10.1093/bioinformatics/btz597
48. Mikheenko A, Kolmogorov M. Assembly Graph Browser: interactive visualization of assembly graphs. Bioinformatics. 2019;35(18):3476-3478. doi: 10.1093/bioinformatics/btz072
49. Wick RR, Schultz MB, Zobel J, Holt KE. Bandage: interactive visualization of de novo genome assemblies. Bioinformatics. 2015;31(20):3350-2. doi: 10.1093/bioinformatics/btv383.
50. Gonnella G, Niehus N, Kurtz S. GfaViz: flexible and interactive visualization of GFA sequence graphs. Bioinformatics. 2019;35(16):2853-2855. doi: 10.1093/bioinformatics/bty1046
51. Rausch T, Snajder R, Leger A, Simovic M, Giurgiu M, Villacorta L, Henssen AG, Fröhling S, Stegle O, Birney E, Bonder MJ, Ernst A, Korbel JO. Long-read sequencing of diagnosis and post-therapy medulloblastoma reveals complex rearrangement patterns and epigenetic signatures. Cell Genom. 2023;3(4):100281. doi: 10.1016/j.xgen.2023.100281.
52. Chen S, et al Paragraph: a graph-based structural variant genotyper for short-read sequence data. Genome Biol. 2019;20(1):291. doi: 10.1186/s13059-019-1909-7.
53. Eggertsson HP, et al GraphTyper2 enables population-scale genotyping of structural variation using pangenome graphs. Nat. Commun. 2019;10:5402.
54. Ebler J, et al. Pangenome-based genome inference allows efficient and accurate genotyping across a wide spectrum of variant classes. Nat Genet. 2022;54(4):518-525. doi: 10.1038/s41588-022-01043-w
55. Vaddadi K, Srinivasan R, Sivadasan N. Read mapping on genome variation graphs. In 19th International Workshop on Algorithms in Bioinformatics (WABI 2019), ed. KT Huber, D Gusfield, art. 7. 2019. Dagstuhl, Ger.: Schloss Dagstuhl–Leibniz-Zent. Inform.
56. Rautiainen M, Marschall T. GraphAligner: rapid and versatile sequence-to-graph alignment. Genome biology 2020;21:1–28.
57. Rajput J, Chandra G, Jain C. Co-linear chaining on pangenome graphs. Algorithms Mol Biol. 2024;19(1):4. doi: 10.1186/s13015-024-00250-w.
58. Sirén J, Välimäki N, Mäkinen V. Indexing Graphs for Path Queries with Applications in Genome Research. IEEE/ACM Trans Comput Biol Bioinform. 2014;11(2):375-88. doi: 10.1109/TCBB.2013.2297101
59. Garrison E, Marth G. Haplotype-based variant detection from short-read sequencing. arXiv preprint arXiv:1207.3907 2012 [q-bio.GN]
60. Poplin R, et al. A universal SNP and small-indel variant caller using deep neural networks. Nat Biotechnol. 2018;36(10):983-987. doi: 10.1038/nbt.4235.

61. Li H, Handsaker B, Wysoker A, Fennell T, Ruan J, Homer N, Marth G, Abecasis G, Durbin R; 1000 Genome Project Data Processing Subgroup. The Sequence Alignment/Map format and SAMtools. Bioinformatics. 2009;25(16):2078-9. doi: 10.1093/bioinformatics/btp352
62. McKenna A, Hanna M, Banks E, Sivachenko A, Cibulskis K, Kernytsky A, Garimella K, Altshuler D, Gabriel S, Daly M, DePristo MA. The Genome Analysis Toolkit: a MapReduce framework for analyzing next-generation DNA sequencing data. Genome Res. 2010;20(9):1297-303. doi: 10.1101/gr.107524.110.
63. Tettelin H, et al Genome analysis of multiple pathogenic isolates of Streptococcus agalactiae: implications for the microbial "pan-genome". Proc Natl Acad Sci U S A. 2005;102(39):13950-5. doi: 10.1073/pnas.0506758102.
64. Decano AG, Downing T. An Escherichia coli ST131 pangenome atlas reveals population structure and evolution across 4,071 isolates. Sci Rep. 2019 9(1):17394. doi: 10.1038/s41598-019-54004-5.
65. Parmigiani L, Wittler R, Stoye J. Revisiting pangenome openness with k-mers 2024;4:e47 Peer Community Journal 10.24072/pcjournal.415
66. Parmigiani L, Garrison E, Stoye J, Marschall T, Doerr D. Panacus: fast and exact pangenome growth and core size estimation. BioRxiv 2024 doi: 10.1101/2024.06.11.598418.
67. Eliseev A, et al Evaluation of haplotype callers for next-generation sequencing of viruses. Infect Genet Evol. 2020;82:104277. doi: 10.1016/j.meegid.2020.104277.
68. Baaijens JA, Stougie L, Schönhuth A. Strain-Aware Assembly of Genomes from Mixed Samples Using Flow Variation Graphs. In: Schwartz, R. (eds) Research in Computational Molecular Biology. RECOMB 2020. Lecture Notes in Computer Science 2020 vol 12074. Springer, Cham. doi: https://doi.org/10.1007/978-3-030-45257-5_14
69. Queiroz VF, et al. Analysis of the Genomic Features and Evolutionary History of Pithovirus-Like Isolates Reveals Two Major Divergent Groups of Viruses. J Virol. 2023;97(7):e0041123. doi: 10.1128/jvi.00411-23.
70. Dziadkiewicz P, Dojer N. Getting insight into the pan-genome structure with PangTree. BMC Genomics. 2020;21(Suppl 2):274. doi: 10.1186/s12864-020-6610-4
71. Wang Z, Jia L, Li J, Liu H, Liu D. Pan-Genomic Analysis of African Swine Fever Virus. Virol Sin. 2020;35(5):662-665. doi: 10.1007/s12250-019-00173-6.
72. Wang L, Luo Y, Zhao Y, Gao GF, Bi Y, Qiu HJ. Comparative genomic analysis reveals an 'open' pan-genome of African swine fever virus. Transbound Emerg Dis. 2020;67(4):1553-1562. doi: 10.1111/tbed.13489.
73. Rodríguez-Pérez R, Fernández L, Marco S. Overoptimism in cross-validation when using partial least squares-discriminant analysis for omics data: a systematic study. Anal Bioanal Chem. 2018;410(23):5981-5992. doi: 10.1007/s00216-018-1217-1
74. Campillo-Balderas JA, Lazcano A, Becerra A. Viral Genome Size Distribution Does not Correlate with the Antiquity of the Host Lineages. Front. Ecol. Evol. 2015;3:143. doi: 10.3389/fevo.2015.00143.
75. Truong QL, Nguyen TL, Nguyen TH, Shi J, Vu HLX, Lai TLH, Nguyen VG. Genome Sequence of a Virulent African Swine Fever Virus Isolated in 2020 from a Domestic Pig in Northern Vietnam. Microbiol Resour Announc. 2021;10(19):e00193-21. doi: 10.1128/MRA.00193-21.
76. Torma G, Tombácz D, Csabai Z, Moldován N, Mészáros I, Zádori Z, Boldogkői Z. Combined Short and Long-Read Sequencing Reveals a Complex Transcriptomic Architecture of African Swine Fever Virus. Viruses. 2021;13(4):579. doi: 10.3390/v13040579.
77. Xiao J, Zhang Z, Wu J, Yu J. A brief review of software tools for pangenomics. Genomics Proteomics Bioinformatics. 2015;13(1):73-6. doi: 10.1016/j.gpb.2015.01.007
78. Downing T, Angelopoulos N. A primer on correlation-based dimension reduction methods for multi-omics analysis. J R Soc Interface. 2023;20(207):20230344. doi: 10.1098/rsif.2023.0344
79. Denti L, et al. Pangenome graph augmentation from unassembled long reads. BioRxiv 2025 doi: https://doi.org/10.1101/2025.02.07.637057
80. Aylward AJ, Petrus S, Mamerto A, Hartwick NT, Michael TP. PanKmer: k-mer-based and reference-free pangenome analysis. Bioinformatics. 2023;39(10):btad621. doi: 10.1093/bioinformatics/btad621

81. Dilthey A, Cox C, Iqbal Z, Nelson MR, McVean G. Improved genome inference in the MHC using a population reference graph. Nat Genet. 2015;47(6):682-8. doi: 10.1038/ng.3257
82. Derelle R, et al Seamless, rapid, and accurate analyses of outbreak genomic data using split k-mer analysis. Genome Res. 2024;34(10):1661-1673. doi: 10.1101/gr.279449.124
83. Hwang S, et al. MEM-based pangenome indexing for k-mer queries. BioRxiv 2024 doi: https://doi.org/10.1101/2024.05.20.595044
84. Pritt J, Chen NC, Langmead B. FORGe: prioritizing variants for graph genomes. Genome Biol. 2018;19(1):220. doi: 10.1186/s13059-018-1595-x
85. Sirén J, et al Personalized Pangenome References. Nat Methods. 2024;21(11):2017-2023. doi: 10.1038/s41592-024-02407-2
86. Tennakoon C, Freville T, Downing T. Panalyze: automated virus pangenome variation graph construction and analysis. BioRxiv 2025 doi: https://doi.org/10.1101/2025.04.10.646565